\documentclass[reprint,amsmath,amssymb,aip,jcp,showpacs,floatfix]{revtex4-1}
\usepackage{graphicx}
\usepackage{dcolumn}
\usepackage{bm}

\usepackage[utf8]{inputenc}
\usepackage[T1]{fontenc}
\usepackage{mathptmx}
\usepackage{multirow}
\usepackage{mathrsfs}
\newcommand\ddfrac[2]{\frac{\displaystyle #1}{\displaystyle #2}}
\begin{document}
\title{On the experimental determination of the repulsive component of the potential from high pressure measurements: what is special about twelve?}
\author{R. Casalini}
\email{riccardo.casalini@nrl.navy.mil}
\author{T. C. Ransom}
\email{timothy.ransom.ctr@nrl.navy.mil}
\altaffiliation{American Society for Engineering Education postdoc}
\affiliation{Naval Research Laboratory, Chemistry Division, Washington DC 20375-5342}
\date{\today}
\begin{abstract}
    In this paper we present an overview of results in the literature regarding the thermodynamical scaling of the dynamics of liquids and polymers as measured from high-pressure measurements. Specifically, we look at the scaling exponent $\gamma$, and argue that it exhibits the limiting behavior $\gamma\rightarrow4$ in regimes for which molecular interactions are dominated by the repulsive part of the intermolecular potential. For repulsive potentials of the form $U(r)\propto r^{-n}$, $\gamma$ has been found to be related to the exponent $n$  via the relation $\gamma=n/3$. Therefore, this limiting behavior for $\gamma$ would suggest that a large number of molecular systems may be described by a common repulsive potential $U(r)\propto r^{-n}$ with $n\approx12$. 
\end{abstract}
\maketitle
The density and temperature dependence of dynamic properties of liquids and polymers (i.e.~viscosity, relaxation and diffusion time) is well described by the thermodynamical scaling (TDS) behavior,\cite{casalini2004thermodynamical,dreyfus2004scaling,alba2004scaling} 
\begin{equation}
    \label{eq:1}
    \log(\textbf{X})=\mathscr{F}(T\,\rho^{-\gamma}),
\end{equation}
where $\textbf{X}$ is a dynamic property $\mathscr{F}$ is an unknown function, $T$ the temperature, $\rho$ the density and $\gamma$ a material dependent parameter. It has been argued that the density scaling of liquid dynamics is better described over a large thermodynamic range substituting $\rho^\gamma$ in Eq.~(\ref{eq:1}) with a density dependent activation energy\cite{alba2002temperature} $E_\infty(\rho)$ and \textit{``that a power-law description of the density dependence of $E_\infty(\rho)$, convenient as it may be, may not carry much physical content."} [\onlinecite{alba2006temperature}]
\begin{figure}[t]
\includegraphics[width=\columnwidth]{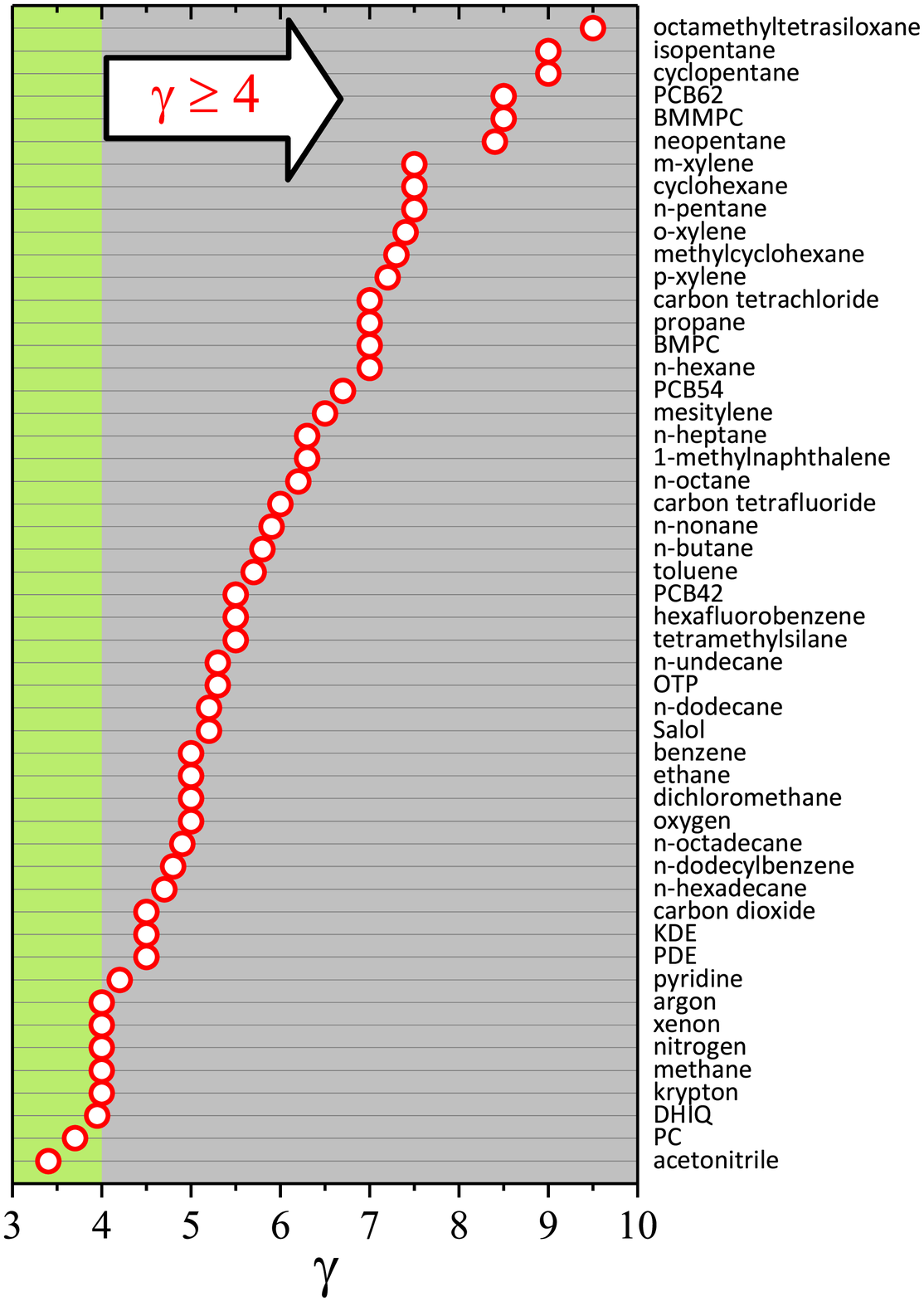}
    \caption{\label{fig:1} Scaling exponent $\gamma$ for fifty-two nonassociated liquids from the literature [\onlinecite{casalini2005liquids,fragiadakis2011connection,abramson2014viscosity,fragiadakis2013polar}].}
\end{figure}

Notwithstanding, disagreements on the use of Eq.~(\ref{eq:1}), the behavior described by the TDS has been verified using different experimental observables for a large number of materials, all with system-dependent values of $\gamma$ which have been found to be generally constant for each material [\onlinecite{roland2005supercooled}]. In particular, for more than fifty nonassociated liquids (Figure \ref{fig:1}), Eq.~(\ref{eq:1}) has been found to describe the dynamics over a large range of pressure and temperature with $\gamma$ for each system lying in the range  $3.5\leq\gamma\leq8.5$\, [\onlinecite{casalini2005liquids,fragiadakis2011connection}]. Probably the more extreme range of pressure over which the TDS behavior has been verified is that of the viscosity measurements of nitrogen for which $\gamma=4$ for up to 10 GPa [\onlinecite{abramson2014viscosity}]. It is noteworthy that of the fifty nonassociated liquids reported in Fig.~\ref{fig:1}, only two have $\gamma<4$, and both liquids are extremely polar, propylene carbonate ($\gamma=3.7$, $\mu\simeq3.9\,D$) and acetonitrile ($\gamma=3.5$, $\mu\simeq4.9\,D$). Molecular dynamics simulations (MDS) have shown that a large dipole moment is expected to cause a decrease of $\gamma$, and therefore the polarity of these two liquids may explain their lower value of $\gamma$ [\onlinecite{fragiadakis2013polar}]. Excluding these two, $\gamma\geq4$ for all other nonassociated liquids. 

Although $\gamma$ has been found to be constant for many systems, some studies have previously reported deviations from the TDS in the form of a changing value of $\gamma$ with density for the liquids Dibutyl phalate (DBP), Decahydroisoquinoline (DHIQ), and DC704,\cite{bohling2012scaling,niss2007correlation,sanz2019experimental}. These findings were anomalous, and subsequent reports showed some inconsistencies in the experimental data of DC704 and DHIQ.\cite{ransom2019comment,casalini2016anomalous} For the case of DBP, it was found\cite{casalini2016density} that there were some inconsistencies with the high pressure viscosity data utilized in Ref.~[\onlinecite{niss2007correlation}], however, even rejecting the viscosity data, an optimal scaling was not found to be possible, and deviation from the scaling was evident in the dielectric data at long times [\onlinecite{casalini2016density}] In particular, the deviation of DBP from the thermodynamical scaling according to Ref..~[\onlinecite{bohling2012scaling}] corresponds to an increase of $\gamma$ with increasing density (and pressure) from 2.6 to 3.9. It is worth nothing that the high pressure behavior of DBP is consistent with that of associated liquids (pressure derivative of the glass temperature, $dT_g/dP=110\,\text{K/GPa}$ and ratio of isochoric and isobaric activation energies, $E_V/E_P=0.7\text{--}0.74$)\cite{sekula2004structural} like tripropylene glycol ($dT_g/dP=109\,\text{K/GPa}$, $E_V/E_P=0.8$),\cite{casalini2003dielectric} and 2-ethyl-1-hexanol ($dT_g/dP=108\,\text{K/GPa}$, $E_V/E_P=0.67$).\cite{fragiadakis2010insights} This is in contrast to the typical high pressure behavior of nonassociated liquids which have been found to have $E_V/E_P\sim\,0.5$ and $dT_g/dP>240\,\text{K/GPa}$.\cite{roland2005supercooled} Thus, in the following we consider DBP as a weakly associated liquid and reanalyze the pressure dependence of the parameter $\gamma$ using a different method than in Ref.~[\onlinecite{bohling2012scaling}]. For a similar reasoning the high pressure behavior of Salol ($dT_g/dP=204\,\text{K/GPa}$ and $E_V/E_P=0.43$)\cite{casalini2003dynamics} is consistent with that of nonassociated liquids and it is included in Fig.~\ref{fig:1}. 

\begin{figure}[t]
\includegraphics[width=\columnwidth]{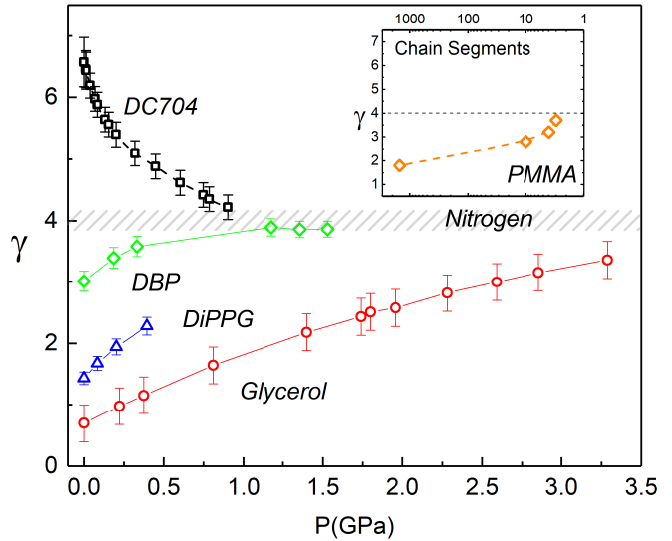}
    \caption{\label{fig:2} Pressure dependence of the scaling exponent $\gamma$ for DC704 [\onlinecite{ransom2019complex}], nitrogen [\onlinecite{abramson2014viscosity}], glycerol [\onlinecite{casalini2011density}], DBP, and DiPPG calculated using Eq.~(\ref{eq:gamma_B}). Inset shows the chain length dependence of $\gamma$ for PMMA [\onlinecite{casalini2007effect}].}
\end{figure}

Recently, an unambiguous deviation from the TDS was observed for the nonassociated liquid DC704, with $\gamma$ decreasing with increasing pressure shown in Figure \ref{fig:2}, from $\gamma=6.6\pm0.4$ at atmospheric pressure to $\gamma=4.2\pm0.4$  at $P=0.9\,\mathrm{GPa}$ [\onlinecite{ransom2019complex}] One particularly interesting part of this behavior is that $\gamma$ appears to decrease rapidly at low pressure, but level off at high pressure close to 4, suggesting that at even higher pressures its value will still be $\gamma\simeq4$. As discussed in greater detail in Ref.~[\onlinecite{ransom2019complex}], an earlier high pressure investigation on DC704 [\onlinecite{gundermann2011predicting}]. did not show this variation of $\gamma$ to be evident because of the limited range of pressure and temperature, even though the range was typical of most investigations found in the literature. Therefore, it is likely that future investigations over similarly broad range of pressure and temperature will evidence more deviations from TDS such that found for DC704 in other simple liquids and some of the values reported previously may be found to represent an average value of $\gamma$.

Summarizing, current experimental results in the literature show that for low to moderately polar, nonassociated liquids: (i) the smallest value of $\gamma$ is $\gamma=4$, (ii) for liquids with $\gamma\simeq4$ at low pressure, no change is observed in $\gamma$ even at 10 GPa, and (iii) for the only case in which a change of $\gamma$ has been observed, an initially large value of $\gamma$ tends to 4 at the highest pressure. This evidence would suggest the possibility of a common limiting value of $\gamma\approx4$ at high densities for simple, nonassociated liquids; so a system with $\gamma\approx4$ at atmospheric pressure will not change very much, but a system with $\gamma>4$ at atmospheric pressure will experience a decrease until it approaches the $\gamma\approx4$ limit.

In order to evaluate the behavior of the parameter $\gamma$ for any system, we briefly
show here a derivation of a simple equation which we then use to calculate the state-point dependence of $\gamma$ using a minimum of inputs.  In Ref.~[\onlinecite{ransom2019complex}], the parameter $\gamma$ of DC704 at various state points was calculated using an approach similar to that used by Sanz et al.\cite{sanz2019experimental}, utilizing the expression
\begin{equation}
    \label{eq:gamma_A}
    \gamma=\frac{\Delta V}{\kappa_TE_P-T\Delta V\alpha_P},
\end{equation}
where $\Delta V=RT(\partial\ln{\textbf{X}}/\partial P)_T$ is the activation volume, $\kappa_T$ the isothermal compressibility, $E_P$ the isobaric activation energy and $\alpha_P$ is the isobaric expansion coefficient.

Using the relationship introduced by Paluch et al. [\onlinecite{paluch2001does}]\footnote{Note that in Ref.~\onlinecite{paluch2001does} Eq.~(\ref{eq:Ep}) is obtained considering that the isobaric fragility $m_T=E_P(RT\,\ln(10))$}
\begin{equation}
    \label{eq:Ep}
    E_P=T\Delta V \frac{\partial P}{\partial T}\Big{|}_\textbf{X},
\end{equation}
Eq.~\ref{eq:gamma_A} can be rewritten as
\begin{equation}
    \label{eq:gamma_B}
    \gamma=\frac{1}{T\Big{(}\kappa_T\ddfrac{\partial P}{\partial T}\Big{|}_\textbf{X}-T\alpha_P\Big{)}}.
\end{equation}
The advantage of this new equation is that only the pressure derivative of the temperature at constant relaxation time (or viscosity (\textbf{X})) is needed together with the pressure and
temperature dependence of $\kappa_T$ and $\alpha_P$ to determine the parameter $\gamma$ at any state point.

Another class of materials for which many confirmations of the TDS behavior with a constant $\gamma$ have been reported is that of polymers. For polymers however, the values of $\gamma$ are generally smaller than for nonassociated liquids, with $1.8\leq\gamma\leq5.6$\, [\onlinecite{casalini2005liquids}]. Larger values of $\gamma$ are associated with polymers having very small torsional potentials (siloxanes) in which the chain constraints are smaller [\onlinecite{casalini2004scaling}]. On the other extreme (of $\gamma$) are polymers with large torsional potential like poly(methylmethacrylate) (PMMA) with $\gamma=1.8$; interestingly for low molecular weight PMMA $\gamma$ increases to $\gamma=3.7$ for the PMMA trimer shown in the inset of Figure \ref{fig:2} [\onlinecite{casalini2007effect}]. Thus the behavior of nonassociated liquids is obtained in the case of polymers in the limits in which the intramolecular potential due to the polymer chain (i.e. torsional potential) is reduced. 

A class of materials for which the TDS has been found to have only limited applicability is hydrogen bonded liquids [\onlinecite{roland2008role,roland2006thermodynamic}]. For these liquids the approximate value of $\gamma$ at low pressure is close to unity. In this study we used Eq.~(\ref{eq:gamma_B}) to calculate the pressure dependence of $\gamma$ for three associated liquids using available dynamic data and EOS data: glycerol (EOS was obtained combining the data from Ref.~[\onlinecite{cibulka1997p}] and Ref.~[\onlinecite{bridgman1926effect}] $(dT/dP)_\tau$ was obtained analyzing the dielectric relaxation data of Ref.~[\onlinecite{johari1972dielectric}], dipropylene glycol (EOS and dielectric relaxation data in Ref.~[\onlinecite{casalini2003dielectric}]) and dibutyl phthalate (EOS from Ref.~[\onlinecite{casalini2016density}] and dielectric relaxation data from Ref.~[\onlinecite{sekula2004structural}]). The pressure behavior of $\gamma$ for these three associated liquids is reported in Fig.~\ref{fig:2} together with those for DC704. Note that the result for DBP is in agreement to that
reported by Bøhling et al. [\onlinecite{bohling2012scaling}]. Differently than for DC704, the three associated liquids
show $\gamma$ increasing with pressure rather than decreasing. For the case of glycerol and
DBP the scaling exponent seems to tend to $\gamma\simeq4$ as shown in figure \ref{fig:2}. Since high pressure decreases the ability of molecules to associate (i.e. hydrogen bond), this observed behavior is suggestive of a high pressure regime in which associated liquids
behaves like a nonassociated liquid with $\gamma\simeq4$.

As discussed above, there are a series of results related with the TDS of associated liquids, nonassociated liquids, and polymers in which a limiting value of $\gamma\simeq4$ is recurrent, which raises the questions: Is there something special about the value $\gamma=4$? Is there a common element to the limiting conditions necessary for the systems to exhibit $\gamma\simeq4$ behavior?

From a theoretical point of view the TDS behavior can be interpreted in terms of the intermolecular potential, $U$ [\onlinecite{casalini2016density,coslovich2008thermodynamic}]. Consider a generalized Lennard-Jones (LJ) potential $U_{LJ}$ [\onlinecite{march2002introduction}],
\begin{equation}
    \label{eq:Potential}
    U_{LJ}=4\epsilon\bigg{[}\bigg{(}\frac{\sigma}{r}\bigg{)}^n-a\bigg{(}\frac{\sigma}{r}\bigg{)}^m\bigg{]},
\end{equation}
where $r$ is the intermolecular distance, $n$ and $m$ are the exponents of the repulsive and attractive terms, $\epsilon$ and $\sigma$ are constants with the dimensions of energy and distance, respectively; while $a$ is constant (equal to either unity or zero) introduced herein to simplify the discussion.

The TDS behavior is strictly predicted for liquids with a purely inverse power law potential (i.e.~$a=0$ in Eq.~(\ref{eq:Potential})) [\onlinecite{hoover1971statistical,hiwatari1974molecular}], with $\gamma=n/3$ . While for liquids in which the attractive term cannot be neglected (i.e.~$a=1$ in Eq.~(\ref{eq:Potential})) the TDS has been found to still be valid using MDS [\onlinecite{coslovich2008thermodynamic,bailey2008pressureI,tsolou2005detailed}] with $\gamma\geq n/3$. This is because the attractive term makes the slope of the potential steeper in the region of the average intermolecular distance, resulting in an effective power law with an exponent seemingly larger than $n$. For conditions of very high pressure and temperature, however, MDS show a decrease of $\gamma$ with increasing density with $\gamma\rightarrow n/3$ in the high pressure limit. It was remarked about this behavior in high pressure conditions, \textit{``this is consistent with the idea that the repulsive part, characterized by an effective inverse power law, dominates the fluctuations"} [\onlinecite{bohling2014estimating}]. Thus, in general for any material taken into consideration, in the high-pressure regime where the applied pressure is so high to render the attractive part of the potential negligible, we would expect that the parameter $\gamma$ will give a measure of the slope exclusively from the repulsive part of the potential. The pressure range to determine this regime evidently depends on the material, with less dense materials (worse packing) expected to show a larger propensity to change, because the pressure can cause a larger change of the average intermolecular distance.  

Since the repulsive part of the intermolecular potential is due to the repulsion of atomic charges, it would not be surprising if the slope of the repulsive part of $U$ could be very similar between different molecules. In principle the repulsive interactions at small intermolecular distance are due to the superposition of electrical repulsion of atomic charges between neutral atoms constituting the molecules. With decreasing intermolecular distance (increasing pressure) the terms related to the closest atoms become dominant (smaller distance may be necessary to overcome the dipole-dipole repulsion in the presence of large dipolar moments). If, in fact, the repulsive part of $U$ is actually very similar for various systems, we would expect that (i) all nonassociated liquids would have $\gamma\geq n/3$ at atmospheric pressure since the attractive part of the potential increases the local slope of the potential at low densities, (ii) for those materials in which $\gamma\simeq n/3$ at low densities (very small attractive term, like for inert gases, i.e.~nitrogen, krypton, argon and xenon\cite{fragiadakis2011connection}) we would expect little or no change of $\gamma$ even at extreme pressure, with the limiting value already attained at very low pressure, and (iii) for those materials having $\gamma$ much larger than $n/3$ we would expect a decrease of $\gamma$ with pressure towards $\gamma\simeq n/3$. Therefore, current experimental results in the literature for simple nonassociated liquids described above are consistent with a common slope of the repulsive part of $U$ that can be approximated by a power law $r^{-n}$ with $n\simeq 12$. Evidently, more experimental
corroboration is needed to confirm the decrease with pressure of the scaling exponent $\gamma$ in other nonassociated liquids, especially those that have been found to have a $\gamma\gg4$ at low pressure. 

The preceding argument proposes that the behavior of non-associated liquids can be explained in terms of a pressure dependent contribution of the attractive part of the potential to the exponent $\gamma$. This contribution becomes progressively smaller with increasing pressure (this can occur at very low pressure in the case of small attractive interactions like for inert gases), but can we use similar arguments to explain the behavior described above for polymers and hydrogen-bonded liquids? For polymers, we noted that where a large torsional potential is present, the value of $\gamma$ is quite a bit lower than 4, but where absent $\gamma$ approaches values close to 4. This torsional potential acts as an intramolecular barrier to rearrangement, reducing the relative influence of intermolecular repulsive potential barriers in regards to molecular relaxation. In the limit of low torsional potential as in the case of siloxanes or the PMMA trimer, the intermolecular potential dominates and the behavior for $\gamma$ becomes like that of nonassociated liquids.  A similar argument can be made for hydrogen-bonded systems, except instead of a torsional potential reducing the relative effect of volume changes, it is hydrogen-bonding networks which compete with and reduce the relative importance of the intermolecular repulsive potential in relation to molecular rearrangements, causing $\gamma<n/3$. With increasing pressure the H-bonding network breaks up causing an increase of $\gamma$ (figure \ref{fig:2}) and at pressures high enough that
the associating potential is overcome by repulsive term, we again obtain the limiting value of $\gamma\approx4$ indicative of interactions dominated by the repulsive potential with the common slope approximated by a $r^{-n}$ power law with $n\simeq12$.

One of the most commonly used potentials in MDS is the 6-12 LJ potential ($n=12$, $m=6$, and $a=1$ in Eq.~(\ref{eq:Potential})). However, to date, the value $n=12$ for the 6-12 LJ potential has never been deducted from first principles or determined experimentally, and it is often used more for convenience than for fundamental reasons. In fact, other forms of $U$ are used in the literature for MDSs with different functional forms of the repulsive term, like for the Buckingham potential [\onlinecite{buckingham1938classical}] where the attractive term is described by an exponential term rather than an inverse power law.

Our review of the data currently available in the literature is consistent with $\gamma\simeq4$ as limit behavior for both associated and nonassociated liquids. However, additional measurements of the dynamic properties and equation of state over a broad range of
temperature and pressure, are necessary to corroborate our observation, especially for
the case of nonassociated liquids having $\gamma\gg4$. Considering the general correlation
found between fragility and $\gamma$, the nonassociated liquids of more interest (i.e.~with the larger $\gamma$) are those with the smaller isochoric fragility [\onlinecite{casalini2005liquids}]. 

In conclusion, we find that the experimental results at high pressure discussed above are consistent with an inverse power law form with $n\simeq12$ as the best approximation of the repulsive term of the intermolecular potential. This does not imply that a 6-12 LJ potential is the best potential to describe the dynamics of nonassociated liquids, since the information in the high-pressure limit is relevant only to small intermolecular distances.   More detailed modeling of the pressure dependence of $\gamma$, would be necessary to extract from high pressure measurements information on other terms in the potential that are relevant for larger intermolecular separation (i.e. attractive terms or repulsive terms of lower order such as dipole-dipole repulsion). We therefore argue that the use of Eq.~(\ref{eq:1}) with a state-point dependent $\gamma$ can be used to extract fundamental information not currently available on the intermolecular potential of liquids. 
\section*{Acknowledgements}
This work was supported by the Office of Naval Research. TCR acknowledges an
American Society for Engineering Education postdoctoral fellowship.
\section*{References}
\providecommand{\noopsort}[1]{}\providecommand{\singleletter}[1]{#1}%

\end{document}